\documentclass[prb,twocolumn,aps,showpacs]{revtex4}
\usepackage{graphicx}%
\usepackage{dcolumn}
\usepackage{amsmath}
\usepackage{amssymb}
\usepackage{epsfig}
\begin{document}

\title{Superfluidity of fermion atoms loaded in a deep optical lattice: the existence of two rotonlike modes }
\author{Z. G. Koinov}\affiliation{Department of Physics and Astronomy,
University of Texas at San Antonio, San Antonio, TX 78249, USA}
\email{Zlatko.Koinov@utsa.edu} 
 \begin{abstract}
We present theoretical calculations of collective modes of  the one-band
attractive Hubbard model which is widely used  to study the s-wave  superfluid phases of
 atomic Fermi gases of two-hyperfine states loaded in a deep optical lattice. To make our
 theory applicable for both superconductivity and superfluidity, we  assume the more general $t-U-J$ Hamiltonian.
  Using the functional differentiation  we derive Schwinger-Dyson equations for the single-particle Green's functions.
  The method of Legendre transform is used to  give a systematic derivation of the Bethe-Salpeter (BS) equation for the
  two-particle Green's function and the associated collective modes. The numerical solution of the BS equation in the limit
  $J\rightarrow 0$ shows the existence of two  rotonlike collective modes with different  low-energy Goldstone dispersions
   and different positions of the rotonlike minima. The two rotonlike modes lie outside of the region determined by the lower
   boundary of the particle-hole continuum, and therefore, the two modes are not damped and  they should be experimentally observable.
   In the presence of superfluid flow at a certain critical flow momentum, the minimum of the first rotonlike mode reaches zero energy,
   but this occurs before the minimum of the second mode and the lower boundary of the particle-hole continuum do, i.e. there are
    two critical flow momenta related to the existence of two rotonlike excitations.
\end{abstract}
 \maketitle
  \section{Introduction}\subsection{The Hamiltonian}
 One of the key
questions of condensed matter physics is to understand the nature
of single particle and collective excitations in high-temperature superconductors.  It is
widely accepted that the Hubbard model with an attractive on-site interaction
 plays an important role in the qualitative understanding
of s-wave superconductors, but there is no consensus, however, about the
theoretical models needed to understand the high-temperature superconductivity. The $t-U-J$ model with a repulsive on-site interaction can serve as a possible model to study  high-temperature superconductivity. This model was
first  used in connection with gossamer
superconductivity,\cite{L,Zh} but it also describes the opening of a d-wave pairing gap in cuprite compounds,
and is  consistent with the fact that the basic pairing
mechanism arises from the antiferromagnetic (AF) exchange
correlations.  On the experimental side one encounters the problem of changing the system parameters $t,U$ and $J$
for a given superconductor, because different parameter
sets require different types of superconducting material.

Ultracold atomic Fermi gases  loaded in optical lattices present a
new opportunity to overcome this obstacle, and to emulate
high-temperature superconductors.  Optical lattices realize the
Hubbard model, if the lattice potential is sufficiently deep such
that the tight-binding approximation is valid. In such a system the
atom-atom interaction can be manipulated in a controllable way by
changing the scattering length from the Bardeen-Cooper-Schrieffer
(BCS) side (negative values) to the Bose-Einstein condensation (BEC)
side (positive values), reaching very large values close to the
Feshbach resonance. On the BEC side of the resonance the
pseudospin-up and pseudospin-down atoms can form diatomic molecules,
and these bosonic molecules can undergo  BEC at a low enough
temperature.\cite{BEC}

In what follows, we  shall focus our attention on the BCS side where
the existence of a superfluid phase of  Fermi atoms is expected
analogous to superconductivity. We shall  examine the spectrum of
the collective excitations of population-balanced atomic Fermi gases
of two hyperfine states with contact interaction, loaded in  optical
lattices. The two hyperfine states are described by pseudospins.
There are $M$ atoms distributed along $N$ sites, and the
corresponding filling factor $f= M / N$ is assumed to be smaller
than unity. For a sufficiently deep lattice potential, the system is
well described by the single-band attractive Hubbard model.

  To make our theory applicable for both superconductivity and superfluidity, the attractive Hubbard model  will be treated as a $J\rightarrow 0$ limit of the more general $t-U-J$ model defined by the following Hamiltonian:
\begin{equation}\begin{split}&H=-\sum_{i,j,\sigma}t_{ij}\psi^\dag_{i,\sigma}\psi_{j,\sigma}
+U\sum_i
\widehat{n}_{i,\uparrow} \widehat{n}_{i,\downarrow}-\mu\sum_{i,\sigma}\widehat{n}_{i,\sigma}\\&+
J\sum_{<i,j>}\overrightarrow{\textbf{S}}_i\textbf{.}\overrightarrow{\textbf{S}}_j\label{H}\end{split}\end{equation} where $t_{ij}$ is the single-electron hopping integral, $\mu$ is the chemical potential, and
 $\widehat{n}_{i,\sigma}=\psi^\dag_{i,\sigma}\psi_{i,\sigma}$ is the
density operator on site $i$. The Fermi operator $\psi^\dag_{i,\sigma}$
($\psi_{i,\sigma}$) creates (destroys) a fermion on the lattice site
$i$  with pseudospin projection $\sigma=\uparrow,\downarrow$. The symbol $\sum_{<ij>}$ means sum over nearest-neighbor sites of
the two-dimensional lattice. The spin operator is defined by  $\overrightarrow{\textbf{S}}_i=(S^x_i,S^y_i,S^z_i)=\psi^\dag_{i,\sigma}
\overrightarrow{\sigma}_{\sigma\sigma'}\psi_{i,\sigma'}/2$, where
$\overrightarrow{\sigma}$ is a vector formed by the Pauli spin
matrices $(\sigma_x, \sigma_y,\sigma_z)$. The AF interaction can be written as $J\sum_{<i,j>}\overrightarrow{\textbf{S}}_i\textbf{.}\overrightarrow{\textbf{S}}_j=
J\sum_{<i,j>}S^z_iS^z_j
+\frac{1}{2}J\sum_{<i,j>}\left[S^+_iS^-_j+S^-_iS^+_j\right]$, where $S^\pm_i=S^x_i\pm
\imath S^y_i$ .

\subsection{Collective excitations in moving optical lattices in the GRPA}
For the case when the periodic array of microtraps is generated by
counter propagating laser beams with differing frequencies the
optical lattice potential is moving with a velocity $-\textbf{v}$
(in the laboratory frame) with magnitude proportional to the
relative frequency detuning of the two laser beams. In a frame fixed
with respect to the lattice potential, the fermion atoms flow with a
constant quasimomentum $\textbf{p}=m\textbf{v}$,  where $m$ is the
mass of the loaded atoms.  For population-balanced  Fermi gases the
order parameter field,
$\Phi_j(u)=-|U|<\psi_{j,\downarrow}(u)\psi_{j,\uparrow}(u)>$ (or
$\Phi^*_j(u)=-|U|<\psi^\dag_{j,\uparrow}(u)\psi^\dag_{j,\downarrow}(u)>$),
in the mean-field approximation varies as
$\Phi_j\propto\Delta\exp\left[2\imath\textbf{p.}\textbf{r}_j\right]$.
Here, the symbol $< >$ means ensemble average, and $\Delta$ is a
real quantity which depends on the lattice velocity.\cite{Gam, Yosh}
In a moving lattice, the formation of  BCS superfluidity is
possible; but due to the presence of quasimomentum $\textbf{p}$, the
superflow can break down.

The stability of balanced superfluid Fermi
gases loaded into a moving optical lattice has been recently studied
using the second-order time-dependent perturbation theory,
\cite{Gam} and the Green's function formalism.\cite{Yosh}  It was pointed out that the
superfluid state could be destabilized at a critical flow momentum
via two different mechanisms: depairing (pair-breaking) at
$\textbf{p}_{pb}$ and Landau instabilities at $\textbf{p}_{cr}$. The
depairing takes place when the single fermionic excitations are
broken, while the Landau instability is related to the rotonlike
structure of the spectrum of the collective excitations. The
superfluid state becomes unstable when the energy of the rotonlike
minimum reaches zero at a given quasimomentum.  The numerical
solution of the number, gap and  collective-mode equations shows
that at a zero temperature the Landau instability appears before the
depairing mechanism.\cite{Yosh}
\subsection{The Hubbard-Stratonovich  transformation and the Bethe-Salpeter equation}

It is known that the single-particle excitations of  the Hamiltonian (\ref{H})   manifest
themselves as poles of the single-particle   Green's function, $G$; while the two-particle
(collective) excitations could be related to
the poles of the  two-particle
Green's function, $K$. The poles of these Green's functions are defined by the solutions of the Schwinger-Dyson (SD)
equation $G^{-1}=G^{(0)-1}-\Sigma$,  and the Bethe-Salpeter (BS)
equation $[K^{(0)-1}-I]\Psi=0$. Here,  $G^{(0)}$ is the
 free single-particle propagator, $\Sigma$ is the electron self-energy,
$I$ is the BS kernel, and the two-particle free propagator $K^{(0)}
= GG $ is a product of two fully dressed single-particle Green's
functions.  Since the electron self-energy
 depends on the two-particle Green's function, the
positions of the poles have to be obtained by solving
 the SD and BS equations self-consistently.

It is widely accepted that the general
random phase approximation (GRPA) is a good approximation for the
collective excitations in a weak-coupling regime, and therefore, it can be used to separate the solutions of the
SD and the BS equations. In this
approximation, the single-particle excitations are replaced with
those obtained by diagonalizing the Hartree-Fock (HF) Hamiltonian; while the
collective modes are obtained by solving the BS equation in which
the single-particle Green's functions are calculated in HF
approximation, and the BS kernel is obtained by summing ladder and
bubble diagrams.

Generally speaking, there exist two different GRPA that can be used to
calculate the spectrum of the collective excitations of the Hubbard Hamiltonian
 in a stationary (or moving) optical lattice. The first approach
uses the Green's function
method,\cite{Yosh,CGexc,CCexc,CG1,ZKexc,Com,ZGK,ZK1}  while the second one is
based on the Anderson-Rickayzen equations.\cite{PA,R,BR,Gam}

The Green's function approach has been used to obtain the collective
excitations in the problem of the exciton BEC
\cite{CGexc,CCexc,ZKexc} and in  s-wave layered superconductors.
\cite{CG1} According to the Green's function method,  the collective
modes  manifest themselves as poles of both the two-particle Green's
function, $K$, and the density and spin response functions. The two
response functions can be expressed in terms of  $K$, but it is very
common to obtain the poles of the density response function by
following the Baym and Kadanoff formalism,\cite{BK} which uses
functional derivatives of the density with respect to the external
fields.\cite{Yosh}

The second method that can be used to obtain the
collective excitation spectrum of the Hubbard Hamiltonian  starts
from the Anderson-Rickayzen equations, which in the GRPA can be
reduced to a set of three coupled equations and the collective-mode
spectrum is obtained by solving  a $3\times 3$ secular determinant. Recently,\cite{Gam} the Belkhir and Randeria $3\times 3$ secular determinant\cite{BR} has been  generalized  to the case
of a moving optical lattice.

 Instead of mean-field decoupling  the quartic interaction
terms, we apply the idea that we can transform these quartic terms
 into  quadratic form by making the
Hubbard-Stratonovich  transformation  for the electron
operators. The attempt to decouple  the quartic $U$ and $J$
terms by using  two-component Nambu field operators $\widehat{\psi}(x)=\left(%
\begin{array}{c}
  \psi_\uparrow(x) \\
    \psi^\dag_\downarrow(x)\\
\end{array}%
\right)$ and $\widehat{\overline{\psi}}
(y)=\left(\psi^\dag_\uparrow(y)\psi_\downarrow(y)
\right)$ does not work because the existence of $S^{\pm}$ parts in the $J$ interaction requires the  use of more complicated  four-component fermion field operators ( $\widehat{\psi}$ and $\widehat{\overline{\psi}}$ obey
anticommutation  relations):
 \begin{equation}\begin{split}&\widehat{\psi}(x)=\frac{1}{\sqrt{2}}\left(%
\begin{array}{c}
  \psi_\uparrow(x) \\
  \psi_\downarrow(x) \\
  \psi^\dag_\uparrow(x)\\
  \psi^\dag_\downarrow(x)\\
\end{array}%
\right),\\&\widehat{\overline{\psi}}
(y)=\frac{1}{\sqrt{2}}\left(\psi^\dag_\uparrow(y)\psi^\dag_\downarrow(y)
\psi_\uparrow(y)\psi_\downarrow(y)
\right).\label{NG1}\end{split}\end{equation}

In contrast to the previous approaches, such that  after performing
the Hubbard-Stratonovich  transformation the fermion degrees of
freedom are integrated out; we decouple the quartic problem by
introducing  a model system which consists of a four-component boson
field $A_{\alpha}(z)$ ($\alpha=1,2,3,4$) interacting with  fermion
fields  (\ref{NG1}).

 There are three
advantages of keeping both the fermion and the boson degrees of
freedom. First,  the approximation that is used to decouple the
self-consistent relation between the electron self-energy and the
two-particle Green's function automatically leads to conserving
approximations because it relies on the fact that the BS kernel
 can be written as functional derivatives of the Fock
$\Sigma^F$ and the Hartree $\Sigma^H$ self-energy
$I=I_d+I_{exc}=\delta \Sigma^F/\delta G+\delta \Sigma^H/\delta
G=\delta^2 \Phi/\delta G\delta G$. As shown by Baym \cite{BK2}, any
self-energy approximation is conserving whenever: (i) the
self-energy  can be written as the derivative of a functional
$\Phi[G]$, i.e. $\Sigma=\delta \Phi[G]/\delta G$, and (ii) the SD
equation for $G$ needs to be solved fully self-consistently for this
form of the self-energy.   Second, the
 collective excitations of the Hubbard model can be calculated in two different ways: as poles of
  the fermion Green's function, $K$, and  as  poles
  of the boson Green's function, $D$; or equivalently,  as poles of the density and spin
parts of the general response function, $\Pi$. Here, the boson
Green's function, $D$, is defined by the equation
  $D=D^{(0)}+D^{(0)}\Pi D^{(0)}$ where $D^{(0)}$ is the free
boson propagator. Third, the action which describes
the interactions in the Hubbard model is similar to the action
$\psi^\dagger A\psi$ in quantum electrodynamics. This allows us to
apply the powerful field-theoretical methods, such as the method of
Legendre transforms, to derive the SD and BS equations, as well as
the
 vertex equation for the vertex function, $\Gamma$,
 and the Dyson equation for the boson Green's function, $D$.

\begin{figure}\includegraphics[scale=0.8]{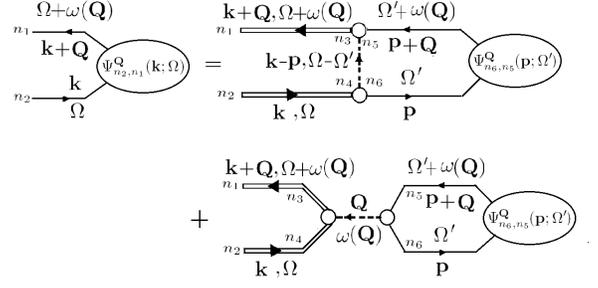}
    \caption{Diagrammatic representation of the Bethe-Salpeter equation (\ref{BSEqq}) for the BS amplitude
    $\Psi^{\textbf{Q}}_{n_2,n_1}(\textbf{k};\Omega)$. The single-particle Green's function $G_{n_2n_1}(\textbf{k},\Omega)$
     is denoted by two solid lines, oriented in the direction of electron propagation. The dashed lines
      represent the free boson propagator $D^{(0)}_{\alpha\beta}(\omega,\textbf{Q})$.
      At each of the bare vertices $\Gamma_\alpha^{(0)}(n_1,n_2)$ (represented by  circles)  the energy and momentum are conserved.}\end{figure}
 The basic assumption in our BS formalism is that the
bound states of two fermions in the optical lattice at zero temperature  are described by the BS wave
functions (BS amplitudes). The BS amplitude determines the probability
amplitude to find the first electron  at the site $i$ at the
moment $t_1$ and the second electron at the site $j$ at the moment $t_2$. The
BS amplitude depends on the relative internal time $t_1-t_2$ and
on the "center-of-mass" time $(t_1+t_2)/2$:\cite{IZ}
\begin{equation}\begin{split}&\Phi^{\textbf{Q}}_{n_2n_1}(\textbf{r}_i,\textbf{r}_j;t_1,t_2)=\\&
\exp\left\{\imath\left[\textbf{Q.}(\textbf{r}_i+\textbf{r}_j)/2-\omega(\textbf{Q})(t_1+t_2)/2\right]\right\}\\&
\Psi^\textbf{Q}_{n_2n_1}(\textbf{r}_i-\textbf{r}_j,t_1-t_2),\end{split}\end{equation}
where $\textbf{Q}$ and $\omega(\textbf{Q})$ are the collective-mode  momentum and the corresponding   dispersion, respectively. Since  $\{n_1, n_2\}=\{1,2,3,4\}$,
we have to take into account the existence of sixteen  BS amplitudes.  The Fourier transform of $\Psi^\textbf{Q}_{n_2n_1}(\textbf{r},t)=\int \frac{d\Omega}{2\pi}\int\frac{d^d\textbf{k}}{(2\pi)^d}e^{\imath\left(\textbf{k}\textbf{.r}-\Omega t\right)}\Psi^{\textbf{Q}}_{n_2,n_1}(\textbf{k};\Omega)$ satisfies the following  BS equation,  presented diagrammatically in FIG. 1:
\begin{equation}\begin{split}
&\Psi^{\textbf{Q}}_{n_2,n_1}(\textbf{k};\Omega)\\&= G_{n_1n_3}\left(\textbf{k}+\textbf{Q},\Omega+\omega(\textbf{Q})\right)G_{n_4n_2}(\textbf{k},\Omega)
\int  \frac{d\Omega'}{2\pi}\int\frac{d^d\textbf{p}}{(2\pi)^d}\\&\left[I_d\left(%
\begin{array}{cc}
  n_3 & n_5  \\
  n_4 & n_6 \\
\end{array}|\textbf{k}-\textbf{p},\Omega-\Omega'
\right)+I_{exc}\left(%
\begin{array}{cc}
  n_3 & n_5  \\
  n_4 & n_6 \\
\end{array}|\textbf{Q},\omega(\textbf{Q})
\right)\right]\\&\Psi^{\textbf{Q}}_{n_6,n_5}(\textbf{p};\Omega').\label{BSEqq}
\end{split}\end{equation}
Here, $G_{n_1n_2}\left(\textbf{k},\Omega\right)$ is the single-particle Green's function, and $I_d$ and $I_{exc}$ are the direct and exchange parts of the BS kernel.

In a similar problem of interacting photons and electrons in electrodynamics, the direct part of the BS kernel does depend on frequency;  therefore  the solution of the BS equation is more complicated. As we shall see in the next Section, in the case of the $t-U-J$ model the boson propagator $D^{(0)}_{\alpha\beta}(\textbf{k})$ ($\alpha,\beta=1, 2, 3, 4$)  is frequency independent, and therefore, the following BS equation for
the equal-time BS amplitude $\Psi^{\textbf{Q}}_{n_2,n_1}(\textbf{k})=\int  \frac{d\Omega}{2\pi}\Psi^{\textbf{Q}}_{n_2,n_1}(\textbf{k};\Omega)$ takes place:
\begin{equation}\begin{split}
&\Psi^{\textbf{Q}}_{n_2,n_1}(\textbf{k})=\int \frac{d\Omega}{2\pi}  G_{n_1n_3}\left(\textbf{k}+\textbf{Q},\Omega+\omega(\textbf{Q})\right)G_{n_4n_2}(\textbf{k},\Omega)\\&\int\frac{d^d\textbf{p}}{(2\pi)^d}
[-\Gamma^{(0)}_\alpha(n_3,n_5)D^{(0)}_{\alpha\beta}(\textbf{k}-\textbf{p})
\Gamma^{(0)}_\beta(n_6,n_4)\\&+
\frac{1}{2}\Gamma^{(0)}_\alpha(n_3,n_4)D^{(0)}_{\alpha\beta}(\textbf{Q})
\Gamma^{(0)}_\beta(n_6,n_5)]\Psi^{\textbf{Q}}_{n_6,n_5}(\textbf{p}).\label{BSEdZ}
\end{split}\end{equation}
In the case of $t-U-J$ model the interaction in the direct kernel can be factorized, i.e. $D^{(0)}_{\alpha\beta}(\textbf{k}-\textbf{p})=f_\alpha(\textbf{k})g_\beta(\textbf{p})$;  therefore it is possible to obtain the collective excitation spectrum  using an $80\times 80$ secular determinant. Such an ambitious
task will be left as a subject of future research. Instead, we shall discuss the limit $J\rightarrow 0$ in which   the dimensions of the secular determinant becomes  $16\times 16$.

The paper is organized as follows. In the next section we apply the functional-integral formalism to derive equations for the single-particle excitations and for the two-particle collective modes. In Section III, we numerically solve the BS equation in the case of 1D and 2D stationary and moving lattices.  We use the mean-field approximation for the single-particle Green's functions, while the spectrum of the  collective excitations is obtained in the GRPA. The technical details have been deferred to the Appendix.

 \section{ FIELD-THEORETICAL APPROACH TO THE COLLECTIVE MODES OF THE  $t-U-J$ MODEL}
\subsection{The functional-integral approach}

  The Green's functions in the  field-theoretical approach are defined by means of the so-called generating functional with sources for the boson and fermion  fields. In our problem  the corresponding functional integrals  cannot be evaluated
exactly because the interaction
part of the Hamiltonian (\ref{H}) is quartic in the Grassmann
fermion fields. However, we can transform the quartic terms to a quadratic
form by introducing  a model system which consists of a
four-component boson field $A_{\alpha}(z)$ ($\alpha=1,2,3,4$)
interacting with fermion fields $\widehat{\overline{\psi}} (y)$
and $\widehat{\psi}(x)$. The spin-dependent nature of the
interactions requires four spin degrees of freedom
  of the Bose
field $A_{\alpha}(z)$. The action of our model system is assumed to
be of the following form
 $S= S^{(e)}_0+S^{(A)}_0+S^{(e-A)}$, where (throughout this paper we use the
summation-integration convention: that repeated variables are summed
up or integrated over):
$$S^{(e)}_0=\widehat{\overline{\psi}
}(y)\widehat{G}^{(0)-1}(y;x)\widehat{\psi} (x),$$
$$S^{(A)}_0=\frac{1}{2}A_{\alpha}(z)D^{(0)-1}_{\alpha
\beta}(z,z')A_{\beta}(z'),$$$$ S^{(e-A)}=\widehat{\overline{\psi}}
(y)\widehat{\Gamma}^{(0)}_{\alpha}(y,x\mid z)\widehat{\psi}
(x)A_{\alpha}(z).$$

 The action $S^{(e)}_0$ describes the fermion (electron) part of the system.
 The inverse Green's function
 of free electrons
$\widehat{G}^{(0)-1}(y;x)$ is $4\times 4$ diagonal matrix:
\begin{equation}\begin{split}
&\widehat{G}^{(0)-1}(y;x)=\\&\sum_{\textbf{k},\omega_m}\exp\left[\imath
\textbf{k.}(\textbf{r}_i-\textbf{r}_{i'})-\omega_m(u-u')\right]
G_{n_1n_2}^{(0)-1}(\textbf{k},\imath\omega_m) , \nonumber
\end{split}\end{equation}
 where  $G_{11}^{(0)-1}(\textbf{k},\imath\omega_m)=
 -G_{33}^{(0)-1}(-\textbf{k},-\imath\omega_m)
 =\imath\omega_m-
\xi(\textbf{k}),$ and
$-G_{22}^{(0)-1}(-\textbf{k},-\imath\omega_m)=G_{44}^{(0)-1}(\textbf{k},\imath\omega_m)=
\imath\omega_m+ \xi(\textbf{k})$.
The symbol $\sum_{\omega_m}$ is used to denote $\beta^{-1}\sum_{m}$
(for fermion fields $\omega_m=
   (2\pi/\beta)(m +1/2) ;m=0, \pm 1, \pm 2,... $), and $\xi(\textbf{k})=2t_x(1-\cos k_x)+2t_y(1-\cos k_y)-\mu$ is the non-interacting dispersion on a square lattice,.

 The action $S^{(A)}_0$ describes
the boson field. The bare boson propagator in $S^{(A)}_0$ provides
the  spin-dependent interactions $U$ and $J$, and it is defined as:
\begin{widetext}
\begin{equation}\widehat{D}^{(0)}
(z,z')=\delta(v-v')\left[U\delta{j,j'}\left(%
\begin{array}{cccc}
  0&1&0&0  \\
 1 &0 &0&0  \\
 0 &0 &0&0\\
 0 &0 & 0&0
\end{array}%
\right)+\sum_{a}\delta_{j',j+a}\left(%
\begin{array}{cccc}
 \frac{1}{2} J-& \frac{1}{2}J &0&0  \\
  \frac{1}{2}J &\frac{1}{2} J &0&0  \\
 0 &0 &0& J  \\
 0 &0 & J&0
\end{array}%
\right)\right].\nonumber\end{equation} Here the summation on $a$
runs over the nearest-neighbor sites of site $j$. The Fourier
transform of the boson propagator is given by
\begin{equation}\begin{split}&
\widehat{D}^{(0)} (z,z')=\frac{1}{N} \sum_\textbf{k}\sum_{\omega_p}
 e^{\left\{\imath\left[\textbf{k.}\left(\textbf{r}_j-\textbf{r}_{j'}\right)
 -\omega_p\left(v-v'\right)\right]\right\}}\widehat{D}^{(0)}(\textbf{k}),\\&
\widehat{D}^{(0)}(\textbf{k})= \left(%
\begin{array}{cccc}
  J(\textbf{k})&U- J(\textbf{k}) &0&0  \\
 U- J(\textbf{k})  &J(\textbf{k}) &0&0  \\
 0 &0 &0&2 J(\textbf{k}) \\
 0 &0 & 2J(\textbf{k})&0
\end{array}%
\right)\label{FTD0}\end{split}\end{equation} where the symbol
$\sum_{\omega_p}$ is used to denote $\beta^{-1}\sum_{p}$ (for boson
fields $\omega_p=
   (2\pi/\beta)p ;p=0, \pm 1, \pm 2,... $), and in the case of a
two-dimensional square lattice  $J(\textbf{k})=J\left(\cos k_x+\cos k_y\right)$.

 The interaction between the fermion and the boson
 fields is described by the action $S^{(e-A)}$.
 The bare vertex
$\widehat{\Gamma}^{(0)}_{\alpha}(y_1;x_2\mid
z)=\widehat{\Gamma}^{(0)}_{\alpha}(i_1,u_1;i_2, u_2\mid
j,v)=\delta(u_1-u_2)\delta(u_1-v)\delta_{i_1i_2}\delta_{i_1j}\widehat{\Gamma}^{(0)}(\alpha)$
is a $4\times 4$ matrix, where
\begin{equation}
\widehat{\Gamma}^{(0)}(\alpha)=\frac{1}{2}(\gamma_0+\alpha_z)\delta_{\alpha1}
+\frac{1}{2}(\gamma_0-\alpha_z)\delta_{\alpha2}+
\frac{1}{2}(\alpha_x+\imath\alpha_y)\delta_{\alpha3}+
\frac{1}{2}(\alpha_x-\imath\alpha_y)\delta_{\alpha4}.\label{Gamma0}\end{equation}
The Dirac matrix $\gamma_0$  and the $4\times 4$ matrices
 $\widehat{\alpha}_i$  are defined as ( $\sigma_i$ are the Pauli matrices):
$$\gamma_0=\left(%
\begin{array}{cccc}
  1&0&0&0  \\
 0&1&0&0  \\
 0& 0& -1&0  \\
 0& 0& 0&-1  \\
\end{array}%
\right),\quad  \widehat{\alpha}_i=\left(%
\begin{array}{cc}
  \sigma_i & 0  \\
 0& \sigma_y\sigma_i\sigma_y \\
\end{array}%
\right), i=x,y,z.$$

The relation between the $t-U-J$  model and our model system can
be demonstrated by applying the Hubbard-Stratonovich transformation
for the electron operators:
\begin{eqnarray}&\int \mu[A]\exp\left[\widehat{\overline{\psi}}
(y)\widehat{\Gamma}^{(0)}_{\alpha}(y;x|z)\widehat{\psi}(x)A_{\alpha}(z)\right]
\nonumber\\&=\exp\left[-\frac{1}{2}\widehat{\overline{\psi}}
(y)\widehat{\Gamma}^{(0)}_{\alpha}(y;x|z)\widehat{\psi}(x)
D_{\alpha,\beta}^{(0)}(z,z') \widehat{\overline{\psi}}
(y')\widehat{\Gamma}^{(0)}_{\beta}(y';x'|z')\widehat{\psi}(x')\right].\label{HSa}\end{eqnarray}
 The functional measure $D\mu[A]$ is chosen to be:
$$
\mu[A]=DAe^{-\frac{1}{2}A_{\alpha}(z)D_{\alpha,\beta}^{(0)-1}(z,z')
A_{\beta}(z')},\int \mu[A] =1.$$

The Hubbard-Stratonovich transformation allows us to map the
$t-U-J$ model onto the model system described by the action $S$.
This transformation creates an extra term that  can be included in the
chemical potential $\overline{\mu}=\mu-fU/2$, and therefore, the mean-field
expression for the chemical potential is recovered by the
Hubbard-Stratonovich transformation. Since there exists a one-to-one
correspondence between the $t-U-J$ model
    and the model system defined by the action  $S$,  we can obtain the
spectrum of the single-particle excitations, as well as the spectrum
of the collective modes by analyzing the single-particle and
two-particle excitations of the model system.

 According to the field-theoretical approach, the expectation value of a general operator
$\widehat{O}(u)$ can be expressed as a functional integral over the
boson field $A$ and the Grassmann fermion fields
$\widehat{\overline{\psi}}$ and $\widehat{\psi}$:
\begin{equation}<\widehat{T}_u(\widehat{O}(u))>=\frac{1}{Z[J,M]}\int
D\mu[\widehat{\overline{\psi}},\widehat{\psi},A]\widehat{O}(u)
\exp\left[J_{\alpha}(z)A_{\alpha}(z)-M(\widehat{\overline{\psi}},\widehat{\psi})\right]|_{J=M=0},\label{Ex}\end{equation}
where the symbol $<...>$ means that the thermodynamic average is
made, and $\widehat{T}_u$ is an $u-$ordering operator. The
functional $Z[J,M]$ is defined by
\begin{equation}
Z[J,M]=\int
D\mu[\widehat{\overline{\psi}},\widehat{\psi},A]\widehat{O}(u)
\exp\left[J_{\alpha}(z)A_{\alpha}(z)-M(\widehat{\overline{\psi}},\widehat{\psi})\right],\label{GW}
\end{equation}where the functional measure
$D\mu[\widehat{\overline{\psi}},\widehat{\psi},A]=DAD\widehat{\overline{\psi}}D\widehat{\psi}
\exp\left(S\right)$ satisfies the condition $\int
D\mu[\widehat{\overline{\psi}},\widehat{\psi},A]=1$. The quantity
$J_\alpha(z)$ is the source of the boson field, while the sources $M_{ij}(y;x)$ of the fermion fields are included in the
$M(\widehat{\overline{\psi}},\widehat{\psi})$ term
:
\begin{eqnarray} &M(\widehat{\overline{\psi}},\widehat{\psi})=
\psi^\dag_\uparrow(y)
M_{11}(y;x)\psi_\uparrow(x)+\psi^\dag_\downarrow(y)
M_{21}(y;x)\psi_\uparrow(x)+\psi^\dag_\uparrow(y) M_{12}(y;
x)\psi_\downarrow(x)+\psi^\dag_\downarrow(y)
M_{22}(y;x)\psi_\downarrow(x)\nonumber\\& + \psi_\uparrow(y)
M_{31}(y;x)\psi_\uparrow(x)+\psi_\downarrow(y)
M_{41}(y;x)\psi_\uparrow(x)+\psi_\uparrow(y)
M_{32}(y;x)\psi_\downarrow(x)+\psi_\downarrow(y)
M_{42}(y;x)\psi_\downarrow(x)\nonumber\\& + \psi^\dag_\uparrow(y)
M_{13}(y;x) x)\psi^\dag_\uparrow(x)+\psi^\dag_\downarrow(y)
M_{23}(y;x)\psi^\dag_\uparrow(x)+\psi^\dag_\uparrow(y)
M_{14}(y;x)\psi^\dag_\downarrow(x)+\psi^\dag_\downarrow(y)
M_{24}(y;x)\psi^\dag_\downarrow(x)\nonumber\\& +\psi_\uparrow(y)
M_{33}(y;x)\psi^\dag_\uparrow(x)+\psi_\downarrow(y)
M_{43}(y;x)\psi^\dag_\uparrow(x)+\psi_\uparrow(y)
M_{34}(y;x)\psi^\dag_\downarrow(x)+\psi_\downarrow(y)
M_{44}(y;x)\psi^\dag_\downarrow(x)
 \label{M1}\end{eqnarray}
Here, we have introduced complex indices $1=\{n_1,x_1\}$, and
$2=\{n_2,y_2\}$ where, $x_1=\{\textbf{r}_{i_1},u_1\}$,
$y_2=\{\textbf{r}_{i_2},u_2\}$ and $\{n_1,n_2\}=\{1,2,3,4\}$. We shall now use  a functional derivative $\delta / \delta
M(2;1)$;  depending on the spin degrees of freedom, there are
sixteen possible derivatives.

 By means of the definition (\ref{Ex}), one can express all Green's functions related
 to system under consideration in terms of the functional derivatives with respect to the
  corresponding  sources
 of the
generating functional of the connected Green's functions $W[J,M]=\ln
Z[J,M]$. Thus, we define the following Green's and vertex functions
which will be used to  analyze the collective modes of our model:

  Boson Green's
function $D_{\alpha \beta}(z,z')$: \\This function is a $4\times 4$
matrix defined as $D_{\alpha \beta}(z,z')=-\frac{\delta^2W}{\delta
J_{\alpha}(z)\delta J_{\beta}(z')}$.

Single-electron Green's function $G_{n_1n_2}(x_1;y_2)$:\\ This
function is a $4\times 4$ matrix whose elements  are
$G_{n_1n_2}(x_1;y_2)=-\delta W/\delta M_{n_2n_1}(y_2;x_1)$:
\begin{eqnarray}&\widehat{G}(x_1;y_2)=\nonumber\\& -\left(%
\begin{array}{cccc}
  <\widehat{T}_u\left(\psi_\uparrow(x_1)\psi^\dag_\uparrow(y_2)\right)> &
<\widehat{T}_u\left(\psi_\uparrow(x_1)\psi^\dag_\downarrow(y_2)\right)>
   & <\widehat{T}_u\left(\psi_\uparrow(x_1)\psi_\uparrow(y_2)\right)> &
     <\widehat{T}_u\left(\psi_\uparrow(x_1)\psi_\downarrow(y_2)\right)> \\
  <\widehat{T}_u\left(\psi_\downarrow(x_1)\psi^\dag_\uparrow(y_2)\right)>
   &
   <\widehat{T}_u\left(\psi_\downarrow(x_1)\psi^\dag_\downarrow(y_2)\right)>
   & <\widehat{T}_u\left(\psi_\downarrow(x_1)\psi_\uparrow(y_2)\right)>&
     <\widehat{T}_u\left(\psi_\downarrow(x_1)\psi_\downarrow(y_2)\right)>\\
   <\widehat{T}_u\left(\psi^\dag_\uparrow(x_1)\psi^\dag_\uparrow(y_2)\right)>
   &
   <\widehat{T}_u\left(\psi^\dag_\uparrow(x_1)\psi^\dag_\downarrow(y_2)\right)>

   & <\widehat{T}_u\left(\psi^\dag_\uparrow(x_1)\psi_\uparrow(y_2)\right)>&
     <\widehat{T}_u\left(\psi^\dag_\uparrow(x_1)\psi_\downarrow(y_2)\right)>\\
   <\widehat{T}_u\left(\psi^\dag_\downarrow(x_1)\psi^\dag_\uparrow(y_2)\right)>
   &<\widehat{T}_u\left(\psi^\dag_\downarrow(x_1)\psi^\dag_\downarrow(y_2)\right)>
   & <\widehat{T}_u\left(\psi^\dag_\downarrow(x_1)\psi_\uparrow(y_2)\right)>
   &
   <\widehat{T}_u\left(\psi^\dag_\downarrow(x_1)\psi_\downarrow(y_2)\right)>  \\
\end{array}%
\right). \label{EGF}
\end{eqnarray}
 Depending on the two spin degrees of freedom,
$\sigma_1$ and $\sigma_2$, there exist eight "normal" Green
functions  and eight "anomalous" Green's functions. We introduce
Fourier transforms of the "normal"
$G_{\sigma_1,\sigma_2}(\textbf{k},u_1-u_2)=
-<\widehat{T}_u(\psi_{\sigma_1,\textbf{k}}(u_1)\psi^\dag_{\sigma_2,\textbf{k}}(u_2))>$,
and "anomalous"
$F_{\sigma_1,\sigma_2}(\textbf{k},u_1-u_2)=-<\widehat{T}_u(\psi_{\sigma_1,\textbf{k}}(u_1)
\psi_{\sigma_2,-\textbf{k}}(u_2))>$ one-particle Green's functions,
where $\{\sigma_1,\sigma_2\}=\uparrow,\downarrow$. Here
$\psi^+_{\uparrow,\textbf{k}}(u),\psi_{\uparrow,\textbf{k}}(u)$ and
$\psi^+_{\downarrow,\textbf{k}}(u),\psi_{\downarrow,\textbf{k}}(u)$
are the creation-annihilation Heisenberg operators. The Fourier
transform of the  single-particle Green's function is given by
\begin{equation}
\widehat{G}(1;2)=
\frac{1}{N}\sum_{\textbf{k}}\sum_{\omega_{m}}\exp\{\imath\left[\textbf{k.}\left(
\textbf{r}_{i_1}-\textbf{r}_{i_2}\right)-\omega_{m}(u_1-u_2)\right]\}
\left(%
\begin{array}{cc}
  \widehat{G}(\textbf{k},\imath\omega_m) & \widehat{F}(\textbf{k},\imath\omega_m)  \\
 \widehat{F}^\dag(\textbf{k},\imath\omega_m) & -\widehat{G}(-\textbf{k},-\imath\omega_m) \\
\end{array}%
\right).\label{FTGF}\end{equation} Here, $\widehat{G}$ and
$\widehat{F}$ are $2\times 2$ matrices whose elements are
$G_{\sigma_1,\sigma_2}$ and $F_{\sigma_1,\sigma_2}$, respectively.

The two-particle
Green's function $K\left(%
\begin{array}{cc}
  n_1,x_1 & n_3,y_3  \\
  n_2,y_2 & n_4,x_4 \\
\end{array}%
\right)$:\\ For different $n_1,n_2,n_3,n_4$, there are $256$
two-particle Green's functions:
\begin{equation}
K\left(%
\begin{array}{cc}
  n_1,x_1 & n_3,y_3  \\
  n_2,y_2 & n_4,x_4 \\
\end{array}%
\right)=K\left(%
\begin{array}{cc}
  1 & 3  \\
  2 & 4 \\
\end{array}%
\right)=\frac{\delta^2 W}{\delta M_{n_2n_1}(y_2;x_1)\delta
M_{n_3n_4}(y_3;x_4)}=-\frac{\delta G_{n_1n_2}(x_1;y_2)}{\delta
M_{n_3n_4}(y_3;x_4)} ; \label{TGF}
\end{equation}
This definition of $K$ allows us to conclude that if the
approximation used for $G$ is chosen in accordance with the recipes proposed by Baym and Kadanoff,\cite{BK2} then $K$ is automatically conserving.

 The vertex function $\widehat{\Gamma}_{\alpha}(2;1 \mid
  z)$:\\ For fixed $\alpha$ the vertex function  is a $4 \times 4$ matrix whose elements are:
\begin{equation}
\widehat{\Gamma}_{\alpha}(i_2,u_2;i_1,u_1 \mid
v,j)_{n_2n_1}=-\frac{\delta G_{n_1n_2}^{-1}(i_2,u_2;i_1,u_1)}{\delta
J_{\beta}(z')} D^{-1}_{\beta \alpha}(z',z). \label{VF}\end{equation}
\subsection{Equations of the boson and fermion Green's functions }
It is well-known that the electron self-energy (electron mass
operator) $\widehat{\Sigma}(1;2)$ can be defined by means of the
so-called SD equations. They can be derived using the fact that the
measure $D\mu[\overline{\psi},\psi,A]$ is invariant under the
translations $\overline{\psi}\rightarrow
\overline{\psi}+\delta\overline{\psi}$ and  $A\rightarrow A+\delta
A$:
\begin{equation}
D^{(0)-1}_{\alpha
\beta}(z,z')R_\beta(z')+\frac{1}{2}Tr\left(\widehat{G}(1;2)\widehat{\Gamma}^{(0)}_{\alpha}(2;1\mid
z)\right)+J_\alpha(z)=0, \label{SE1}
\end{equation}
\begin{equation}
\widehat{G}^{-1}(1;2)-\widehat{G}^{(0)-1}(1;2)+\widehat{\Sigma}(1;2)+\widehat{M}(1;2)=0,
\label{SE2}
\end{equation}
where $R_\alpha(z)=\delta W/\delta J_\alpha(z)$ is the average boson
field. The electron self-energy   $\widehat{\Sigma}$ is a $4\times
4$ matrix which can be written as a sum of Hartree
$\widehat{\Sigma}^H$ and Fock $\widehat{\Sigma}^F$ parts. The
Hartree part is a diagonal matrix whose elements are:
\begin{equation}
\Sigma^H(i_1,u_1;i_2,u_2)_{n_1n_2}=\frac{1}{2}
\widehat{\Gamma}_{\alpha}^{(0)}(i_1,u_1;i_2,u_2|j,v)_{n_1n_2}
D^{(0)}_{\alpha\beta}(j,v;j',v')\widehat{\Gamma}_{\beta}^{(0)}(i_3,u_3;i_4,u_4|j',v')_{n_3n_4}
G_{n_4n_3}(i_4,u_4;i_3,u_3)
.\label{H1}
\end{equation}

The Fock part of the electron self-energy is given by:
\begin{equation}\begin{split}&
\Sigma^F(i_1,u_1;i_2,u_2)_{n_1n_2}=-
\widehat{\Gamma}_{\alpha}^{(0)}(i_1,u_1;i_6,u_6|j,v)_{n_1n_6}
D^{(0)}_{\alpha\beta}(j,v;j',v')\widehat{\Gamma}_{\beta}^{(0)}(i_4,u_4;i_5,u_5|j',v')_{n_4n_5}
\times\\&K\left(%
\begin{array}{cc}
  n_5,i_5,u_5 & n_3,i_3,u_3  \\
  n_4,i_4,u_4 & n_6,i_6,u_6 \\
\end{array}%
\right)G^{-1}_{n_3n_2}(i_3,u_3;i_2,u_2).\label{Fock
sigma}\end{split}
\end{equation}
The Fock part of the electron self-energy depends on the
two-particle Green's function $K$, and therefore, the SD equations
and the BS equation for $K$ have to be solved self-consistently.

Our approach to the $t-U-J$ model allows us to obtain exact
equations of the Green's functions by using the field-theoretical
technique. We now wish to return to our statement that the Green's
functions are the thermodynamic average of the
$\widehat{T}_u$-ordered products of field operators. The standard
procedure for calculating the Green's functions is to apply Wick's
theorem. This  enables us to evaluate the $\widehat{T}_u$-ordered
products of field operators as a perturbation expansion involving
only wholly contracted field operators. These expansions can be
summed formally to yield different equations of Green's functions.
The main disadvantage of this procedure is that the validity of the
equations must be verified diagram by diagram. For this reason we
will  use the method of Legendre transforms of the generating
functional for connected Green functions.\cite{DM} By applying the
same steps as  in Ref. [\onlinecite{ZKexc}] we obtain the BS
equation of the two-particle Green's function, the Dyson equation of
the boson Green's function, and the vertex equation:
\begin{equation}
 K^{-1}\left(%
\begin{array}{cc}
  n_2,i_2,u_2 & n_3,i_3,u_3  \\
  n_1,i_1,u_1 & n_4,i_4,u_4 \\
\end{array}%
\right)= K^{(0)-1}\left(%
\begin{array}{cc}
  n_2,i_2,u_2 & n_3,i_3,u_3  \\
  n_1,i_1,u_1 & n_4,i_4,u_4 \\
\end{array}%
\right)-I\left(%
\begin{array}{cc}
  n_2,i_2,u_2 & n_3,i_3,u_3  \\
  n_1,i_1,u_1 & n_4,i_4,u_4 \\
\end{array}%
\right),\label{BSK}
\end{equation}
\begin{equation}
D_{\alpha \beta}(z,z')=D^{(0)}_{\alpha \beta}(z,z')+D^{(0)}_{\alpha
\gamma}(z,z'')\Pi_{\gamma\delta}(z'',z''')D^{(0)}_{\delta
\beta}(z,z'),\label{BosonDyson}\end{equation}
\begin{equation}\begin{split}
&\widehat{\Gamma}_{\alpha}(i_2,u_2;i_1,u_1\mid
z)_{n_2n_1}=\widehat{\Gamma}^{(0)}_{\alpha}(i_2,u_2;i_1,u_1\mid
z)_{n_2n_1}+I\left(%
\begin{array}{cc}
 n_2,i_2,u_2 & n_3,i_3,u_3  \\
  n_1,i_1,u_1 & n_4,i_4,u_4 \\
\end{array}%
\right)\times\\&
K^{(0)}\left(%
\begin{array}{cc}
 n_3,i_3,u_3 & n_6,i_6,u_6  \\
  n_4,i_4,u_4 & n_5,i_5,u_5 \\
\end{array}%
\right)\widehat{\Gamma}_{\alpha}(i_6,u_6;i_5,u_5\mid z)_{n_6n_5}.
\label{Edward}\end{split}\end{equation}
Here, $$K^{(0)}\left(%
\begin{array}{cc}
 n_2,i_2,u_2 & n_3,i_3,u_3  \\
  n_1,i_1,u_1 & n_4,i_4,u_4 \\
\end{array}%
\right)=G_{n_2n_3}(i_2,u_3;i_2,u_2)G_{n_4n_1}(i_4,u_4;i_1,u_1)$$  is
the two-particle free propagator constructed from a pair of fully
dressed single-particle Green's functions. The kernel
$I=\delta\Sigma/\delta G$ of the BS equation can be expressed as a
functional derivative  of the electron self-energy
$\widehat{\Sigma}$. Since
$\widehat{\Sigma}=\widehat{\Sigma}^H+\widehat{\Sigma}^F$, the BS
kernel $I=I_{exc}+I_d$ is a sum of
 functional derivatives of the Hartree $\Sigma_H$ and
Fock $\Sigma_F$ contributions to the self-energy:
\begin{equation}
I_{exc}\left(%
\begin{array}{cc}
 n_2,i_2,u_2 & n_3,i_3,u_3  \\
  n_1,i_1,u_1 & n_4,i_4,u_4 \\
\end{array}%
\right)=\frac{\delta\Sigma^H(i_2,u_2;i_3,u_3)_{n_2n_1}}{\delta
G_{n_3n_4}(i_3,u_3;i_4,u_4)},\quad
I_d\left(%
\begin{array}{cc}
 n_2,i_2,u_2 & n_3,i_3,u_3  \\
  n_1,i_1,u_1 & n_4,i_4,u_4 \\
\end{array}%
\right)=\frac{\delta\Sigma^F(i_2,u_2;i_3,u_3)_{n_2n_1}}{\delta
G_{n_3n_4}(i_3,u_3;i_4,u_4)}.\label{Kernel}\end{equation} The
general response function $\Pi$ in the Dyson equation (\ref{BosonDyson}) is defined as
\begin{equation}\Pi_{\alpha \beta}(z,z')=
\widehat{\Gamma}^{(0)}_{\alpha}(i_1,u_1;i_2,u_2 \mid
z)_{n_1n_2}K\left(%
\begin{array}{cc}
  n_2,i_2,u_2 & n_3,i_3,u_3  \\
  n_1,i_1,u_1 & n_4,i_4,u_4 \\
\end{array}%
\right)\widehat{\Gamma}^{(0)}_{\beta}(i_3,u_3,i_4,u_4\mid
z')_{n_3n_4} .\label{Pi}
\end{equation}
 The  functions $D$, $K$  and $\widehat{\Gamma}$
 are related by the  identity:
\begin{equation}\begin{split}&
K^{(0)}\left(%
\begin{array}{cc}
 n_2,i_2,u_2 & n_3,i_3,u_3  \\
  n_1,i_1,u_1 & n_4,i_4,u_4 \\
\end{array}%
\right)\widehat{\Gamma}_{\beta}(i_4,u_4;i_3,u_3\mid
z')_{n_4n_3}D_{\beta \alpha}(z',z)\\&= K\left(%
\begin{array}{cc}
  n_2,i_2,u_2 & n_3,i_3,u_3  \\
  n_1,i_1,u_1 & n_4,i_4,u_4 \\
\end{array}%
\right)\widehat{\Gamma}^{(0)}_{\beta}(i_4,u_4;i_3,u_3\mid
z')_{n_4n_3}D^{(0)}_{\beta \alpha}(z',z), \label{GammaK}
\end{split}\end{equation}

 By introducing the boson proper self-energy $P^{-1}_{\alpha
\beta}(z,z')=
\Pi^{-1}_{\alpha\beta}(z,z')+D^{(0)}_{\alpha\beta}(z,z')$ one can
rewrite the Dyson equation (\ref{BosonDyson})
 for the boson Green's function as:
\begin{equation}D^{-1}_{\alpha \beta}(z,z') =D^{(0)-1}_{\alpha
\beta}(z,z')-P_{\alpha \beta}(z,z').\label{DE}\end{equation} The
proper self-energy and the vertex function $\widehat{\Gamma}$ are
related by the following equation:
 \begin{equation}\begin{split}&
P_{\alpha
\beta}(z,z')=\frac{1}{2}Tr\left[\widehat{\Gamma}_\alpha^{(0)}(y_1,x_2|z)\widehat{G}(x_2,y_3)
\widehat{\Gamma}_\beta(y_3,x_4|z')\widehat{G}(x_4,y_1)\right]\\&=\frac{1}{2}
\widehat{\Gamma}^{(0)}_{\alpha}(i_1,u_1;i_2,u_2\mid
z)_{n_1n_2}G_{n_2n_3}(i_2,u_2;i_3,u_3)
\widehat{\Gamma}_{\beta}(i_3,u_3;i_4,u_4\mid
z')_{n_3n_4}G_{n_4n_1}(i_4,u_4;i_1,u_1).
\label{PG}\end{split}\end{equation} It is also possible to express the
proper self-energy in terms of the two-particle Green's function
$\widetilde{K}$ which satisfies the BS equation
$\widetilde{K}^{-1}=K^{(0)-1}-I_d$, but its kernel
$I_d=\delta\Sigma^F/\delta G$ includes only diagrams that represent
the direct interactions:
  \begin{equation}\begin{split}&
P_{\alpha
\beta}(z,z')=\widehat{\Gamma}^{(0)}_{\alpha}(i_1,u_1;i_2,u_2\mid
z)_{n_1n_2}\widetilde{K}\left(%
\begin{array}{cc}
  n_2,i_2,u_2 & n_3,i_3,u_3  \\
  n_1,i_1,u_1 & n_4,i_4,u_4 \\
\end{array}%
\right) \widehat{\Gamma}^{(0)}_{\beta}(i_3,u_3;i_4,u_4\mid
z')_{n_3n_4}\\&
=\widehat{\Gamma}^{(0)}_{n_1n_2}(\alpha)\widetilde{K}\left(%
\begin{array}{cc}
  n_2,\textbf{r}_j,v & n_3,\textbf{r}_{j'},v'  \\
  n_1,\textbf{r}_j,v & n_4,\textbf{r}_{j'},v' \\
\end{array}\right)\widehat{\Gamma}^{(0)}_{n_3n_4}(\beta). \label{P}\end{split}\end{equation}
 The proper boson self-energy $\widehat{P}$, the vertex function $\widehat{\Gamma}$ and
the two-particle Green's function $\widetilde{K}$ have common poles.
Let $\omega_{l\textbf{Q}}$ and $\textbf{Q}$ denote the energy and
momentum of one of these common poles. Close to $\omega_{l\textbf{Q}}$ one
can write:
\begin{equation}
\widetilde{K}\left(%
\begin{array}{cc}
  n_1,i_1,u_1 & n_3,i_3,u_3  \\
 n_2,i_2,u_2 & n_4,i_4,u_4 \\
\end{array}%
\right)\approx \sum_{\omega_{p}}e^{-\imath
\omega_{p}(u_1-u_3)}\frac{\psi_{n_2,n_1}^{l\textbf{Q}}(\textbf{r}_{i_2},\textbf{r}_{i_1};u_2-u_1)
\psi_{n_3,n_4}^{l\textbf{Q} \ast
}(\textbf{r}_{i_3},\textbf{r}_{i_4};u_4-u_3)}{\imath \omega_p -
\omega_{l\textbf{Q}}},\label{Gf}\end{equation} where the amplitudes
$\psi_{n_2,n_1}^{l\textbf{Q}}(\textbf{r}_{i_2},\textbf{r}_{i_1};u_2-u_1)$
have the following form:
$$\psi_{n_2,n_1}^{l\textbf{Q}}(\textbf{r}_{i_2},\textbf{r}_{i_1};u_2-u_1)=\exp
\left[\imath \textbf{Q.}\left(\textbf{r}_{i_1}+\textbf{r}_{i_1}\right)/2\right]
\psi^{l\textbf{Q}}_{n_2,n_1}(\textbf{r}_{i_2}-\textbf{r}_{i_1};u_2-u_1).$$
Due to the definition of the bare vertex $\widehat{\Gamma}^{(0)}$,
we have to take into account only the equal "time" $u_1=u_2$
amplitudes:
$$\psi^{l\textbf{Q}}_{n_2,n_1}(\textbf{r}_{i_2}-\textbf{r}_{i_1};0)=
\frac{1}{N}\sum_{\textbf{k}} \exp\{\imath\textbf{k.}
\left(\textbf{r}_{i_2}-\textbf{r}_{i_1}\right)\}\psi^{l\textbf{Q}}_{n_2,n_1}(\textbf{k}),$$
where $\psi^{l\textbf{Q}}_{n_2,n_1}(\textbf{k})$ is the equal "time"
two-particle wave functions in $\textbf{k}$-representation.  By
means of (\ref{Gf}) and (\ref{P}) we obtain:
\begin{equation}
P_{\alpha\beta}(\textbf{Q},\omega)=\sum_{l}\left[
\frac{\varphi^{l\textbf{Q}}_\alpha\varphi^{\ast
l\textbf{Q}}_\beta}{\omega-\omega_{l\textbf{Q}}+\imath 0^+}
-\frac{\varphi^{\ast
l\textbf{Q}}_\alpha\varphi^{l\textbf{Q}}_\beta}{\omega+\omega_{l\textbf{Q}}+\imath
0^+} \right],\label{PO}
\end{equation}
where $\omega_{l\textbf{Q}}=E_{l}(\textbf{Q})-\overline{\mu}$, and
$\varphi^{l\textbf{Q}}_\alpha=\delta_{\alpha 1}\left(
\psi^{l\textbf{Q}}_{1,1}(0;0)-\psi^{l\textbf{Q}}_{3,3}(0;0)\right)
+\delta_{\alpha 2}\left(
\psi^{l\textbf{Q}}_{2,2}(0;0)-\psi^{l\textbf{Q}}_{4,4}(0;0)\right)+\delta_{\alpha
3}\left(
\psi^{l\textbf{Q}}_{1,2}(0;0)-\psi^{l\textbf{Q}}_{4,3}(0;0)\right)+\delta_{\alpha
4}\left(
\psi^{l\textbf{Q}}_{2,1}(0;0)-\psi^{l\textbf{Q}}_{3,4}(0;0)\right)$.
The Green's function (\ref{Gf}) is invariant under the exchange
$4\leftrightarrow 1$ and $2\leftrightarrow 3$, therefore,
$\varphi^{l\textbf{Q}}_1=-\varphi^{l\textbf{Q}}_{2}$, and
$\varphi^{l\textbf{Q}}_3=-\varphi^{l\textbf{Q}}_{4}$. Thus, we
obtain that $P_{11}=P_{22}$, and $P_{33}=P_{44}$. This means that
the proper self-energy has a maximum of six different elements:
\begin{equation}P_{\alpha\beta}(\textbf{Q};\omega)=\left(%
\begin{array}{cccc}
  P_{11}(\textbf{Q};\omega)&P_{12}(\textbf{Q};\omega)&P_{13}(\textbf{Q};\omega)&P_{14}(\textbf{Q};\omega)\\
  P_{12}(\textbf{Q};\omega)&P_{11}(\textbf{Q};\omega)&P_{14}(\textbf{Q};\omega)&P_{13}(\textbf{Q};\omega)\\
    P_{13}(\textbf{Q};\omega)&P_{14}(\textbf{Q};\omega)&P_{33}(\textbf{Q};\omega)&P_{34}(\textbf{Q};\omega)\\
      P_{14}(\textbf{Q};\omega)&P_{13}(\textbf{Q};\omega)&P_{34}(\textbf{Q};\omega)&P_{33}(\textbf{Q};\omega)\\
\end{array}%
\right)=\left(%
\begin{array}{cccc}
  a&b&c&d\\
  b&a&d&c\\
    c&d&f&h\\
      d&c&h&f\\
\end{array}%
\right).\label{Palphabeta}\end{equation}

 \subsection{Spin-singlet  order parameter}
  As we have already mentioned, the BS equation and
the SD equations  have to be solved self-consistently. The so-called
$D^{(0)}\Gamma^{(0)}$ approximation allows us to decouple the
above-mentioned equations and to obtain a linearized integral
equation for the Fock term. To apply  this approximation we first
use Eq. (\ref{GammaK}) to rewrite the Fock term as
\begin{equation}\Sigma^F(i_1,u_1;i_2,u_2)_{n_1n_2}=-
\widehat{\Gamma}_{\alpha}^{(0)}(i_1,u_1;i_3,u_3|j,v)_{n_1n_3}
D_{\alpha\beta}(j,v;j',v')G_{n_3n_4}(i_3,u_3;i_4,u_4)\widehat{\Gamma}_{\beta}(i_4,u_4;i_2,u_2|j',v')_{n_4n_2}
,\label{MassFock}\end{equation} and after that we replace $D$ and
$\widehat{\Gamma}$ in (\ref{MassFock}) by the free boson propagator
$D^{(0)}$  and by the bare vertex $\widehat{\Gamma}^{(0)}$,
respectively.  In this approximation the Fock term assumes the form:
\begin{equation}
\Sigma_0^F(i_1,u_1;i_2,u_2)_{n_1n_2}=-
\widehat{\Gamma}_{\alpha}^{(0)}(i_1,u_1;i_3,u_3|j,v)_{n_1n_3}
D^{(0)}_{\alpha\beta}(j,v;j',v')\widehat{\Gamma}_{\beta}^{(0)}(i_4,u_4;i_2,u_2|j',v'))_{n_4n_2}
G_{n_3n_4}(i_3,u_3;i_4,u_4)\label{SF1}
\end{equation}
In this approximation the total self-energy is defined as\cite{HT}
\begin{equation}\widehat{\Sigma}(i_1,u_1;i_2,u_2)=
-U\delta_{i_1,i_2}\delta(u_1-u_2)\left(%
\begin{array}{cccc}
  -G_{22}(1;2) &G_{12}(1;2)&0&-G_{14}(1;2)  \\
 G_{21}(1;2)& -G_{11}(1;2) &-G_{23}(1;2)&0  \\
 0 & -G_{32}(1;2)&-G_{44}(1;2)&G_{34}(1;2)  \\
 -G_{41}(1;2) & 0&G_{43}(1;2)&-G_{33}(1;2)  \\
\end{array}%
\right). \label{HFG}\end{equation}
The contributions to
$\Sigma(i_1,u_1;i_2,u_2)$ due to the elements on the major
diagonal of the above matrices will be included into the chemical
potential $\overline{\mu}$. To obtain analytical expression for the single-particle Green's function, we will
assume two more approximations. First, we neglect
$G_{12}=G_{21}=G_{34}=G_{43}=0$, and second, we neglect the
frequency dependence of the Fourier transform of the Fock part of
the electron self-energy , i.e.
$\widehat{\Sigma}^F(\textbf{k},\imath\omega_m)\approx\widehat{\Delta}(\textbf{k})$,
where the order parameter $\widehat{\Delta}$ is a $4\times 4$
matrix. It is known that in order to preserve the antisymmetry in
the case of a spin-singlet Cooper-pairing, the order parameter
$\Delta(\textbf{k}) $ must be an even function of $\textbf{k}$, so
the momentum dependence of $\widehat{\Sigma}^F(\textbf{k})$ will be

\begin{equation}
\widehat{\Delta}(\textbf{k})=\left(%
\begin{array}{cccc}
 0 &0&0&\Delta(\textbf{k})  \\
 0& 0 &-\Delta(\textbf{k})&0  \\
 0 &-\Delta(\textbf{k})&0&0  \\
 \Delta(\textbf{k}) & 0&0&0   \\
\end{array}%
\right).\nonumber\end{equation}  In this approximation the Fourier
transform of the single-particle Green's function assumes the
following form:
\begin{equation}
\widehat{G}(\textbf{k},\imath\omega_m)=
\frac{1}{(\imath\omega_m)^2-E^2(\textbf{k})}\left(\begin{array}{cccc}
 \imath\omega_m+\xi(\textbf{k}) &0&0&\Delta(\textbf{k})  \\
 0&\imath\omega_m+\xi(\textbf{k})& -\Delta(\textbf{k}) & 0  \\
 0 &-\Delta(\textbf{k})&\imath\omega_m-\xi(\textbf{k}) &0  \\
 \Delta(\textbf{k}) & 0&0&\imath\omega_m-\xi(\textbf{k})    \\
\end{array}%
\right),\label{GFS}
\end{equation} where
$E(\textbf{k})=\sqrt{\xi^2(\textbf{k})+\Delta(\textbf{k})^2}$
and
$\xi(\textbf{k})=2t_x(1-\cos k_x)+2t_y(1-\cos k_y)-\overline{\mu}$.
Substituting this single-particle  Green's function in Eq.   (\ref{SF1}) we reobtain the  gap
equation:
\end{widetext}
\begin{equation}\begin{split}&\Delta(\textbf{k})=
\frac{1}{N}\sum_\textbf{q}
\left[-U+3J(\textbf{k}-\textbf{q})\right]\frac{\Delta(\textbf{q})}{2E(\textbf{q})}\tanh\left(\frac{\beta
E(\textbf{q})}{2}\right)
.\label{Sgap}\end{split}\end{equation}

In the case of the s-wave superfluidity  we have $J\rightarrow 0$,
$\Delta(\textbf{k})=\Delta$ and $U<0$, and by using  the above
single-particle Green's function we obtain the  gap equation for
$\Delta$ and the particle number equation for the filling factor:
\begin{equation}\begin{split}&
f=\frac{2}{N}\sum_\textbf{k} \left[u^2(\textbf{k})
f\left(E(\textbf{k})\right)+
v^2(\textbf{k})f\left(-E(\textbf{k})\right)\right],
\\& 1=\frac{|U|}{N}\sum_\textbf{k}
\frac{f\left(-E(\textbf{k})\right)-
f\left(E(\textbf{k})\right)}{2E_{\textbf{p}}(\textbf{k})},
\label{MO}\end{split}\end{equation} $f(x)=\left[\exp\left(\beta x\right)+1\right]^{-1}$ is the Fermi
distribution function, and the coherence factors are  $u^2(\textbf{k})=\left[1+
\xi(\textbf{k})/E(\textbf{k})\right]/2$,
$v^2(\textbf{k})=1-u^2(\textbf{k})$.

The  gap  equation and the particle number equation in the case of a moving lattice are given in the Appendix.
 \subsection{ Collective-mode equations}

It is easy to see that in the case of a spin-singlet pairing $a-b=h$
and  $c=d=f=0$. From the Dyson equation (\ref{DE}), it follows that the
spectrum of the collective excitations $\omega(\textbf{Q})$ could be
obtained by setting  the $4\times 4$ determinant
$det|D_{\alpha\beta}^{(0)-1}(\textbf{Q};\omega)-P_{\alpha\beta}(\textbf{Q};\omega)|=0$,
or
$det|\delta_{\alpha\beta}-D_{\alpha\gamma}^{(0)}(\textbf{Q};\omega)
P_{\gamma\beta}(\textbf{Q};\omega)|=0$. Thus, we obtain the following exact equations for the collective
modes:
\begin{equation}
0=1+\left[U-2J(\textbf{Q})\right](a-b), \label{FSS1}
\end{equation}
\begin{equation}
0=1-U(a+b), \label{FSS2}
\end{equation}
\begin{equation}
0=1-2J(\textbf{Q})(a-b), \label{FSS3}
\end{equation}
Since the  equations (\ref{BosonDyson}) and (\ref{Pi}) hold,
 the Fourier transforms of the general response function
(\ref{Pi}), the two-particle Green's function $K$, as well as the
Fourier transform of the boson $D_{\alpha\beta}$ Green's function
should share the poles defined by (\ref{FSS1}) - (\ref{FSS3}).

To separate the spin and charge contributions to the poles of the
general response function  we express
$\Pi_{\alpha\beta}(\textbf{Q};\omega)$ in terms of the Fourier
transform of the proper self-energy
$P_{\alpha\beta}(\textbf{Q};\omega)$:
\begin{equation}
\Pi^{-1}_{\alpha\beta}(\textbf{Q};\omega)=P^{-1}_{\alpha\beta}(\textbf{Q};\omega)
-D^{(0)}_{\alpha\beta}(\textbf{Q};\omega). \label{GRF}\end{equation}

By means of (\ref{GRF}) we obtain
$$\widehat{\Pi}(\textbf{Q};\omega)=
\widehat{\Pi}^c(\textbf{Q};\omega)+
\widehat{\Pi}^s(\textbf{Q};\omega),$$ where
$\Pi^s_{\alpha\beta}(\textbf{Q};\omega)$ and
$\Pi^c_{\alpha\beta}(\textbf{Q};\omega)$ represent the contributions
to the general response function due to the spin and  charge
fluctuations:
\begin{equation}\begin{split}&\widehat{\Pi}^s=\\&
\left(%
\begin{array}{cccc}\Pi^{(1)}_{s}(\textbf{Q};\omega)&-\Pi^{(1)}_{s}(\textbf{Q};\omega)&0&0\\
  -\Pi^{(1)}_{s}(\textbf{Q};\omega)& \Pi^{(1)}_{s}(\textbf{Q};\omega)&0&0\\
    0&0&0&\Pi^{(2)}_{s}(\textbf{Q};\omega)\\
      0&0&\Pi^{(2)}_{s}(\textbf{Q};\omega)&0\\
\end{array}%
\right),\label{PiSSinglet}\end{split}\end{equation}
\begin{equation}\widehat{\Pi}^c=
\left(%
\begin{array}{cccc}\Pi_c(\textbf{Q};\omega)&\Pi_c(\textbf{Q};\omega)&0&0\\
 \Pi_c(\textbf{Q};\omega)& \Pi_c(\textbf{Q};\omega)&0&0\\
    0&0&0&0\\
      0&0&0&0\\
\end{array}%
\right),\label{PiDSinglet}\end{equation} where
\begin{equation}\begin{split}
&\Pi^{(1)}_{s}(\textbf{Q};\omega)=\frac{(a-b)/2}{1+\left[U-2J(\textbf{Q})\right](a-b)},\\&
\Pi^{(2)}_{s}(\textbf{Q};\omega)=\frac{a-b}{1-2J(\textbf{Q})(a-b)},\\&
\Pi_c(\textbf{Q};\omega)=\frac{(a+b)/2}
{1+U(a+b)}.\label{Pisd}\end{split}\end{equation}
The collective modes originating from spin fluctuations manifest
themselves as poles of the following  spin response functions (spin susceptibilities):
\begin{equation}\begin{split}&\chi_1^s(\textbf{Q};\omega)=2\Pi^{(1)}_{s}(\textbf{Q};\omega)
=\frac{\kappa^s(\textbf{Q};\omega)}
{1+[U-2J(\textbf{Q})]\kappa^s(\textbf{Q};\omega)},\\&
\chi_2^s(\textbf{Q};\omega)=\Pi^{(2)}_{s}(\textbf{Q};\omega)
=\frac{\kappa^s(\textbf{Q};\omega)}
{1-2J(\textbf{Q})\kappa^s(\textbf{Q};\omega)}
\label{J1Singlet}\end{split}\end{equation}
 where
$\kappa^s(\textbf{Q};\omega) =a-b$.  The second collective mode,
defined by Eq. (\ref{FSS2}), manifests itself as a pole of the
charge density (charge
susceptibility):
\begin{equation}\chi^c(\textbf{Q};\omega)=2\Pi_c(\textbf{Q};\omega)=
\frac{\kappa^{c}(\textbf{Q};\omega)}
{1+U\kappa^{c}(\textbf{Q};\omega)},
\label{DSinglet}\end{equation}  where $\kappa^{c}(\textbf{Q};\omega)
=a+b$.

\subsection{ Random phase approximation }
The collective modes can be obtained from   Eqs. (\ref{FSS1}) -
(\ref{FSS3}) by  solving  the
 BS equation for the two-particle Green's function  $\widetilde{K}$ . The solutions of the BS equation are not
needed in the RPA, in which $\widetilde{K}\simeq K^{(0)}$; or,
equivalently,  in Eq. (\ref{PG}) for the proper self-energy  we replace
 $\widehat{\Gamma}$  by $\widehat{\Gamma}^{(0)}$:
\begin{equation}\begin{split}&P^{RPA}_{\alpha\beta}(z;z')=\\&\frac{1}{2}Tr\left[\widehat{\Gamma}_\alpha^{(0)}(y,x|z)\widehat{G}(x,y')
\widehat{\Gamma}_\beta^{(0)}(y',x'|z')\widehat{G}(x',y)\right].\label{RPAS}\end{split}\end{equation}

The RPA expression for the proper self-energy provides
$a-b=h=\kappa^{(0)}_{zz}(\textbf{Q};\imath
\omega_{p})=\kappa^{(0)}_{xx}(\textbf{Q};\imath
\omega_{p})=\kappa^{(0)}_{yy}(\textbf{Q};\imath
\omega_{p})=\kappa_{0}^{s}(\textbf{Q};\imath \omega_{p})$,
$a+b=\kappa_{0}^{c}(\textbf{Q};\imath \omega_{p})$ and $c=d=f=0$,
where the general  susceptibility
$\kappa^{(0)}_{ij}(\textbf{Q};\imath \omega_{p}), i,j=x,y,z$ was
introduced decades ago:\cite{B}
\begin{equation}\begin{split}&\kappa^{(0)}_{ij}(\textbf{Q};\imath
\omega_{p})=\\&
\frac{1}{2}\sum_{\omega_{m}}\sum_\textbf{k}Tr\left\{\widehat{\alpha}_i
\widehat{G}(\textbf{k};\imath \omega_{m})\widehat{\alpha}_j
\widehat{G}(\textbf{k}+\textbf{Q};\imath \omega_{p}+\imath
\omega_{m})\right\}. \label{RPA1}
 \end{split}
 \end{equation}

 The noninteracting charge and spin
susceptibilities $\kappa_{0}^{s}(\textbf{Q};\omega)$
and $\kappa_{0}^{c}(\textbf{Q};\omega)$ are:
\begin{equation}\begin{split}&
\left\{%
\begin{array}{cccc}
  \kappa_0^s(\textbf{q},\imath\omega_p)\\
 \kappa_0^c(\textbf{q},\imath\omega_p)
\end{array}%
\right\}=\\&\frac{1}{N}\sum_\textbf{k}\left\{%
\begin{array}{c}
  \gamma^2_{\textbf{k},\textbf{q}}\\
 l^2_{\textbf{k},\textbf{q}}
\end{array}%
\right\}\frac{\zeta(\textbf{k},\textbf{q})
[f(E(\textbf{k}))-f(E({\textbf{k}+\textbf{q}))}]}
{(\imath\omega_p)^2-\zeta^2(\textbf{k},\textbf{q})}\\&
+\frac{1}{N}\sum_\textbf{k}\left\{%
\begin{array}{cccc}
  \widetilde{\gamma}^2_{\textbf{k},\textbf{q}}\\
 m^2_{\textbf{k},\textbf{q}}
\end{array}%
\right\}\frac{\varepsilon(\textbf{k},\textbf{q})
[1-f(E(\textbf{k}))-f(E({\textbf{k}+\textbf{q}))}]}
{(\imath\omega_p)^2-\varepsilon^2(\textbf{k},\textbf{q})}.
\label{RPA}\end{split}\end{equation} Here,
$\zeta(\textbf{k},\textbf{q})=
 E(\textbf{k}+\textbf{q})- E(\textbf{k})$, $\varepsilon(\textbf{k},\textbf{q})=
 E(\textbf{k}+\textbf{q})+ E(\textbf{k})$, and the following  form factors have been used:
\begin{equation}\begin{split}&\gamma_{\textbf{k},\textbf{q}}=
u_{\textbf{k}}u_{\textbf{k}+\textbf{q}}
+v_{\textbf{k}}v_{\textbf{k}+\textbf{q}},\quad
l_{\textbf{k},\textbf{q}}=u_{\textbf{k}} u_{\textbf{k}+\textbf{q}}
-v_{\textbf{k}}v_{\textbf{k}+\textbf{q}},\\&
\widetilde{\gamma}_{\textbf{k},\textbf{q}}=
u_{\textbf{k}}v_{\textbf{k}+\textbf{q}}
-u_{\textbf{k}+\textbf{q}}v_{\textbf{k}},\quad
 m_{\textbf{k},\textbf{q}}=
u_{\textbf{k}}v_{\textbf{k}+\textbf{q}}+
u_{\textbf{k}+\textbf{q}}v_{\textbf{k}}.
\label{FF1}
\end{split}\end{equation}
\begin{figure}\includegraphics[scale=0.6]{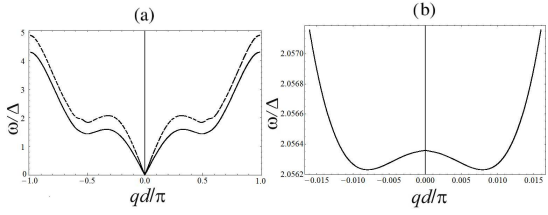}
    \caption{Excitation spectra in 1D stationary optical lattice.
    We set $t_x=t$, $f=0.5$, $U=2t$. The superfluid gap and the chemical potential are
$\Delta=0.409 t$ and $\mu=0.624 t$. The solid  line in (a)
represents the spectrum obtained by solving the equation
$F_1(\omega,q)=0$, while the dashed line is plotted using
$F_2(\omega,q)=0$. The spectrum (b) represents the solution of the
equation $F_3(\omega,q)=0$.}\end{figure}
\begin{figure}\includegraphics[scale=0.6]{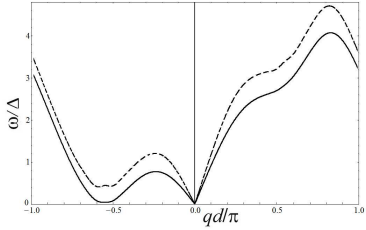}
    \caption{Excitation spectra in 1D moving optical lattice (p=0.21).
    We set $t_x=t$, $f=0.5$, $U=2t$. The superfluid gap and the chemical potential
    are $\Delta=0.420 t$ and $\mu=0.655 t$. The solid  line
represents the spectrum obtained by solving the equation
$F_1(\omega,q)=0$, while the dashed line is plotted using
$F_2(\omega,q)=0$.}\end{figure}
\begin{figure}\includegraphics[scale=0.9]{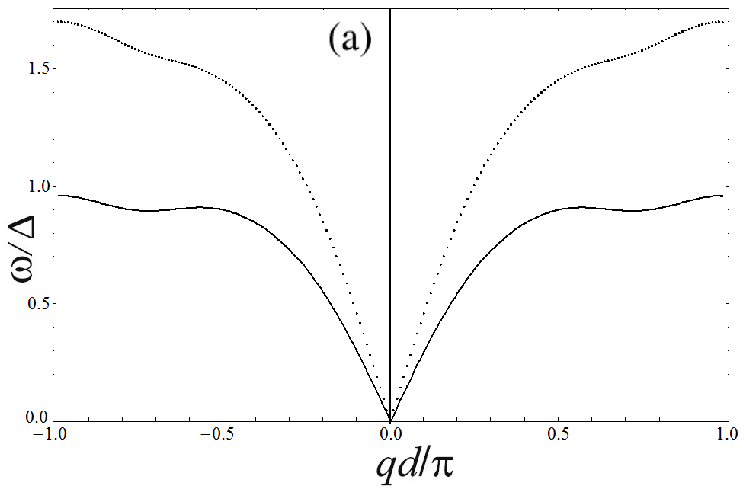}\\\includegraphics[scale=0.9]{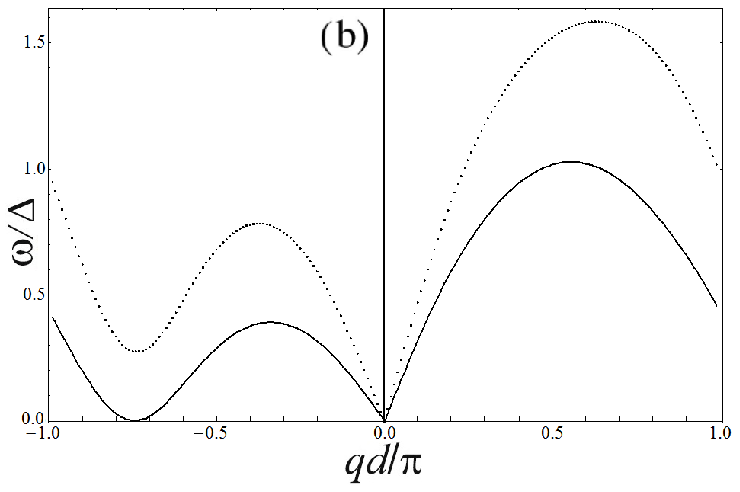}
    \caption{Excitation spectra in 2D optical lattices for  stationary (a) and moving (b) ($p_x=p_y=0.467/\sqrt{2}$) lattices.  We set $t_x=t_y=t$, $f=0.5$, $U=4.5t$. The superfluid gap and chemical potential in the case of a stationary lattice  are
$\Delta=1.334 t$ and $\mu=2.234 t$. For a moving lattice  we have
$\Delta=1.386 t$ and $\mu=2.287 t$. The solid  lines in (a) and (b)
represent the spectrum obtained by solving the equation
$F_1(\omega,q)=0$, while the dotted lines are plotted using
$F_2(\omega,q)=0$. }\end{figure}
\section{ Collective modes of  stationary and moving optical lattices at  zero temperature}
In the case of  deep optical lattices we neglect the AF interaction,
and we assume an attractive Hubbard interaction. The collective
spectrum is obtained by solving the  BS equation at  zero
temperature in the GRPA. In this approximation the corresponding
equation for the BS amplitude  $\Psi^{\textbf{Q}}_{n_2,n_1}=\int
\frac{d\Omega}{2\pi}\int
\frac{d^d\textbf{k}}{(2\pi)^d}\Psi^{\textbf{Q}}_{n_2,n_1}(\textbf{k};\Omega)$
can be obtained from Eq. (\ref{BSEdZ}) by  performing  integration
over the momentum vectors:
\begin{equation}\begin{split}
&\Psi^{\textbf{Q}}_{n_2n_1}=K^{(0)}\left(%
\begin{array}{cc}
  n_1 & n_3  \\
  n_2 & n_4 \\
\end{array}%
|\omega(\textbf{Q})\right)\times\\&
\left[I_d\left(%
\begin{array}{cc}
  n_3 & n_5  \\
  n_4 & n_6 \\
\end{array}%
\right)+I_{exc}\left(%
\begin{array}{cc}
  n_3 & n_5  \\
  n_4 & n_6 \\
\end{array}%
\right)\right]\Psi^{\textbf{Q}}_{n_6,n_5},\label{BSEdZ1}
\end{split}\end{equation}
where the two-particle propagator $K^{(0)}$ and  the direct and exchange interactions are defined  as follows:
\begin{equation}\begin{split}&K^{(0)}\left(%
\begin{array}{cc}
  n_1 & n_3  \\
  n_2 & n_4 \\
\end{array}%
|\omega(\textbf{Q})\right)\equiv K_{n_1n_3n_4n_2}=\\&\int \frac{d\Omega}{2\pi} \int\frac{d^d\textbf{k}}{(2\pi)^d}G_{n_1n_3}\left(\textbf{k}+\textbf{Q},\Omega+\omega(\textbf{Q})\right)G_{n_4n_2}(\textbf{k},\Omega),\\
&I_d\left(%
\begin{array}{cc}
  n_1 & n_3  \\
  n_2 & n_4 \\
\end{array}%
\right)=-\Gamma^{(0)}_\alpha(n_1,n_3)D^{(0)}_{\alpha\beta}
\Gamma^{(0)}_\beta(n_4,n_2),
\\&I_d\left(%
\begin{array}{cc}
  n_1 & n_3  \\
  n_2 & n_4 \\
\end{array}%
\right)=
\frac{1}{2}\Gamma^{(0)}_\alpha(n_1,n_2)D^{(0)}_{\alpha\beta}
\Gamma^{(0)}_\beta(n_4,n_3).\label{BSEdZ2}
\end{split}\end{equation}
The boson propagator  $D^{(0)}_{\alpha\beta}$ in (\ref{BSEdZ2}) is given by (\ref{FTD0}) in the limit $J(\textbf{k})=0$.

The non-trivial solution of the BS equations (\ref{BSEdZ1}) exists if the   secular determinate $det|\widehat{I}+U\widehat{Z}|$ is zero, where $\widehat{I}$ is the unit matrix, and the $16 \times 16$ matrix $\widehat{Z}$ is given in the Appendix. After computing the secular determinant, we find that  at a given point $\textbf{Q}$ there exist three different dispersions $\omega_j(\textbf{Q}), j=1,2,3$, which are the solutions of the three equations $F_j(\omega_j,\textbf{Q})=0$, where:
\begin{widetext}
\begin{equation}\begin{split}&
F_1(\omega,\textbf{Q})=1+\left(I_{l,l}+I_{m,m}+I_{\gamma,\gamma}\right)U+\left(-I^2_{l,m}+I_{l,l}I_{m,m}-J^2_{\gamma,l}-I^2_{\gamma,m}+
I_{l,l}I_{\gamma,\gamma}+I_{m,m}I_{\gamma,\gamma}\right)U^2+\\&
\left(-I_{m,m}J^2_{\gamma,l}+2I_{l,m}J_{\gamma,l}I_{\gamma,m}-I_{l,l}I^2_{\gamma,m}-I_{\gamma,\gamma}I^2_{l,m}+I_{l,l}I_{m,m}I_{\gamma,\gamma}\right)U^3
\label{F1}\end{split}\end{equation}
\begin{equation}\begin{split}&
F_2(\omega,\textbf{Q})=1+\left(-I_{m,m}+2I_{\gamma,\gamma}\right)U+\left(-I^2_{l,l}+J^2_{l,\widetilde{\gamma}}-I_{l,l}I_{m,m}
-J^2_{\gamma,l}+I_{l,l}I_{\gamma,\gamma}-I_{m,m}I_{\gamma,\gamma}+I^2_{\gamma,\gamma}+I^2_{\gamma,\widetilde{\gamma}}\right)U^2+\\&
\left(I_{l,l}J^2_{\gamma,l}+I_{m,m}J^2_{\gamma,l}-I_{\gamma,\gamma}I^2_{l,l}+I_{\gamma,\gamma}J^2_{l,\widetilde{\gamma}}-I_{\gamma,\gamma}I_{l,l}I_{m,m}
-I_{\gamma,\gamma}J^2_{\gamma,l}+I_{l,l}I^2_{\gamma,\gamma}-2J_{l,\widetilde{\gamma}}J_{\gamma,l}I_{\gamma,\widetilde{\gamma}}+
I_{l,l}I^2_{\gamma,\widetilde{\gamma}}\right)U^3
\label{F2}\end{split}\end{equation}
\begin{equation}
F_3(\omega,\textbf{Q})=1+\left(-I_{l,l}-2I_{m,m}+I_{\gamma,\gamma}\right)U+\left(I_{l,l}I_{m,m}+I^2_{m,m}-J^2_{\widetilde{\gamma},m}-
I_{\gamma,\gamma}I_{m,m}\right)U^2
\label{F3}\end{equation}
\end{widetext}
The definitions of $I_{a,b}$ and $J_{a,b}$ are given in the Appendix, and we have used the relation $I_{\widetilde{\gamma},\widetilde{\gamma}}=I_{l,l}+I_{m,m}-I_{\gamma,\gamma}$.

The  dispersion of the first rotonlike mode $\omega_1(\textbf{Q})$, obtained by  solving numerically  the equation $F_1(\omega_1,\textbf{Q})=0$  has been previously obtained by the equation-of-motion method,\cite{Gam} and by locating the poles of the density response function.\cite{Yosh}  The numerical solution of the equation $F_2(\omega_2,\textbf{Q})=0$ provides  the existence of a second  rotonlike collective mode $\omega_2(\textbf{Q})$. The two rotonlike modes have  different  low-energy Goldstone dispersions  and different  rotonlike minima.

 In  FIG. 2 - FIG. 4  we have plotted the excitation spectrum of superfluid Fermi gases
 in 1D and 2D optical lattices assuming the same strength of the interactions  and filling factors
  as in FIG. 4 ($U=2t, f=0.5$) and as in FIG. 9 ($U=4.5t, f=0.5$) in Ref. [\onlinecite{Yosh}]. In the presence of
superfluid flow, as the flow momentum increases, the  two rotonlike
spectra lean toward the left side and the energies of the two
rotonlike minima  decrease.  It is worth mentioning that the two
rotonlike modes lie outside of the region determined by the lower
boundary of the particle-hole continuum;  therefore the two modes
are not damped and constitute propagating modes and they should be
experimentally observable. At a certain flow momentum, the minimum
of the first rotonlike mode reaches zero energy, but this occurs
before  both the  minimum of the second mode, and the lower boundary
of the particle-hole continuum do. At this critical flow momentum
the first mode  is  destabilized  due to the spontaneous emission of
rotonlike excitations; but to   destabilize the second mode one has
to increase the flow momentum. Thus, there exist two critical
lattice velocities which are  determined by the existence of two
rotonlike modes.

The third mode, $\omega_3(\textbf{Q})$, is obtained by solving the equation  $F_3(\omega_3,\textbf{Q})=0$. This mode  lies entirely inside the particle-hole continuum Therefore,  it is a non-propagating mode and cannot be an experimentally observable mode.
\section{Conclusion }
In conclusion, we have  studied the superfluid state of  Fermi gases
in deep optical lattices. Using the Bethe-Salpeter equation in the
GRPA we have obtained the collective mode spectrum of the attractive
Hubbard model in the presence of superfluid flow. We found that the
spectrum of the collective excitations has two rotonlike modes with
different  low-energy Goldstone dispersions  and different rotonlike
minima. Both modes are  experimentally observable because they are
separated from the region determined by the lower boundary of the
particle-hole continuum. As the flow momentum increases, the two
rotonlike spectra lean toward the left side. At a certain critical
flow momentum, the energy minimum of the first collective mode  hits
zero, but this occurs before both the  minimum of the second mode,
and the lower boundary of the particle-hole continuum do.

 The  collective mode spectrum of the $t-U$ model can be obtained by applying the Kadanoff-Baym method for constructing the linear response function of the system.  The linear response function of the system can be obtained by using  four component fermion fields and a  single-particle Green's function represented by a four by four matrix.\cite{HT} In this case
 the Kadanoff-Baym method leads to a four by four  secular determinant which provides  only one roton mode. As can be seen, the single-particle Green's function and the electron self-energy in the present work and in  Ref. [\onlinecite{HT}]  are exactly the same,  but the BS approach and  the  Kadanoff-Baym method provide different secular determinants. In the case when the  single-particle Green's function is a four by four matrix, the BS amplitude is a column matrix with sixteen rows and  the corresponding secular determinant  is a sixteen by sixteen matrix.
 The  result obtained by the  Kadanoff-Baym method can be derived within the BS approach by  keeping in the kernel of the BS equation only diagrams that will reduce the BS amplitude from a sixteen by one matrix to a column matrix with four rows.\cite{ZRM}

The important question is the physical nature of the two gapless collective modes. These Nambu-Goldstone (NG) modes should correspond to the spontaneously broken internal symmetry generators: one quasiparticle with no energy gap for each spontaneously broken symmetry. In the case of superfluidity, both the particle number symmetry and Galilean symmetry are spontaneously broken. Unfortunately, the number of NG modes associated with the spontaneous breaking that appear in a nonrelativistic system is a complicated problem.  It was pointed out by Nielsen and Chadha\cite{NC} that the NG modes are of two types: in the case of type-I the energy $\omega(\textbf{Q})$ is odd powers of $\textbf{Q}$ while in the type-II we have even powers of  $\textbf{Q}$.  The number of type-I modes
plus twice the number of type-II modes is greater than or equal to the number of broken symmetries. As we have pointed out, there are only two NG modes of type-I, and therefore, our BS approach is in accordance with the  Nielsen and Chadha theorem.

\begin{widetext}
\newpage
\textbf{Appendix}

The BS equation (\ref{BSEdZ1}) written in the matrix form is
$\left(\widehat{I}+U\widehat{Z}\right)\widehat{\Psi}=0$, where
$\widehat{I}$ is the  $16 \times 16$ unit matrix. The    $16 \times
16$ matrix $\widehat{Z}$ is defined as follows:
\begin{equation}
\widehat{Z}=\left(%
\begin{array}{cccccccccccccccc}
  -\frac{K_{1414}}{2}&0&0&K_{1411}&0&\frac{K_{1111}}{2}&0&0&0&0&
  \frac{K_{1414}}{2}&0&K_{1114}&0&0&-\frac{K_{1111}}{2}\\
  0&-K_{1111}&0&0&0&0&0&0&0&0&0&0&0&0&K_{1411}&0\\
  0&K_{1411}&0&0&0&0&0&0&0&0&0&0&0&0&-K_{4414}&0\\
 -\frac{K_{4414}}{2} &0&0&K_{4411}&0&\frac{K_{1411}}{2}&0&0&0&0&
 \frac{K_{4414}}{2}&0&K_{1414}&0&0&-\frac{K_{1411}}{2}\\
 0&0&0&0&-K_{1111}&0&0&0&0&0&0&K_{1414}&0&0&0&0\\
 \frac{K_{1111}}{2}&0&0&0&0&-\frac{K_{1414}}{2}&-K_{1411}&0&0&
 -K_{1114}&-\frac{K_{1111}}{2}&0&0&0&0& \frac{K_{1414}}{2}\\
 -\frac{K_{1411}}{2}&0&0&0&0&\frac{K_{4414}}{2}&K_{4411}&0&0&K_{1414}&\frac{K_{1411}}{2}&0&0&0&0&-\frac{K_{4414}}{2}\\
 0&0&0&0&-K_{1411}&0&0&0&0&0&0&K_{4414}&0&0&0&0\\
 0&0&0&0&K_{1114}&0&0&0&0&0&0&-K_{1444}&0&0&0&0\\
 -\frac{K_{1114}}{2}&0&0&0&0&\frac{K_{1444}}{2}&K_{1414}&0&0&K_{1144}&\frac{K_{1114}}{2}&0&0&0&0&-\frac{K_{1444}}{2}\\
 \frac{K_{1414}}{2}&0&0&0&0&-\frac{K_{4444}}{2}&-K_{4414}&0&0&-K_{1444}&-\frac{K_{1414}}{2}&0&0&0&0&\frac{K_{4444}}{2}\\
 0&0&0&0&K_{1414}&0&0&0&0&0&0&-K_{4444}&0&0&0&0\\
 -\frac{K_{1444}}{2}&0&0&K_{1414}&0&\frac{K_{1114}}{2}&0&0&0&0&\frac{K_{1444}}{2}&0&K_{1144}&0&0&-\frac{K_{1114}}{2}\\
 0&-K_{1114}&0&0&0&0&0&0&0&0&0&0&0&0&K_{1444}&0\\
 0&K_{1414}&0&0&0&0&0&0&0&0&0&0&0&0&-K_{4444}&0\\
 -\frac{K_{4444}}{2}&0&0&K_{4414}&0&\frac{K_{1414}}{2}&0&0&0&0&\frac{K_{4444}}{2}&0&K_{1444}&0&0&-\frac{K_{1414}}{2}\\
\end{array}%
\right)\nonumber\end{equation}
The transposed matrix of $\widehat{\Psi}$  is given by:
$$\widehat{\Psi}^T=
\left(%
\begin{array}{cccccccccccccccc}
  \Psi^{\textbf{Q}}_{1,1} &
  \Psi^{\textbf{Q}}_{1,2} &
  \Psi^{\textbf{Q}}_{1,3} &
  \Psi^{\textbf{Q}}_{1,4} &
  \Psi^{\textbf{Q}}_{2,1} &
  \Psi^{\textbf{Q}}_{2,2} &
  \Psi^{\textbf{Q}}_{2,3} &
  \Psi^{\textbf{Q}}_{2,4} &
  \Psi^{\textbf{Q}}_{3,1} &
  \Psi^{\textbf{Q}}_{3,2} &
  \Psi^{\textbf{Q}}_{3,3} &
  \Psi^{\textbf{Q}}_{3,4} &
  \Psi^{\textbf{Q}}_{4,1} &
  \Psi^{\textbf{Q}}_{4,2} &
  \Psi^{\textbf{Q}}_{4,3} &
  \Psi^{\textbf{Q}}_{4,4} \\
\end{array}%
\right).
$$
At  zero temperature the elements of matrix $\widehat{Z}$ are:
\begin{equation}\begin{split}
&K_{1111}=\frac{1}{2}\left(I_{m,m}+I_{\widetilde{\gamma},\widetilde{\gamma}}-2J_{\widetilde{\gamma}, m}\right),\quad K_{4444}=\frac{1}{2}\left(I_{m,m}+I_{\widetilde{\gamma},\widetilde{\gamma}}+2J_{\widetilde{\gamma}, m}\right),\quad K_{1144}=\frac{1}{2}\left(I_{l,l}+I_{\gamma,\gamma}+2J_{\gamma, l}\right)\\&
K_{4411}=\frac{1}{2}\left(I_{l,l}+I_{\gamma,\gamma}-2J_{\gamma, l}\right),\quad
K_{1114}=\frac{1}{2}\left(-I_{l,m}+I_{\gamma,\widetilde{\gamma}}+J_{l,\widetilde{\gamma}}-J_{\gamma,m}\right),\quad
K_{4414}=\frac{1}{2}\left(I_{l,m}-I_{\gamma,\widetilde{\gamma}}+J_{l,\widetilde{\gamma}}-J_{\gamma,m}\right)\\&
K_{1414}=\frac{1}{2}\left(I_{l,l}-I_{\gamma,\gamma}\right),\quad K_{1444}=\frac{1}{2}\left(I_{l,m}+I_{\gamma,\widetilde{\gamma}}+J_{l,\widetilde{\gamma}}+J_{\gamma,m}\right),\quad
K_{1411}=\frac{1}{2}\left(-I_{l,m}-I_{\gamma,\widetilde{\gamma}}+J_{l,\widetilde{\gamma}}-J_{\gamma,m}\right),
\nonumber\end{split}\end{equation}
where the following symbols are used:
$$I_{a,b}=\frac{1}{N}\sum_\textbf{k}
\frac{a_{\textbf{k},\textbf{Q}}b_{\textbf{k},\textbf{Q}}\varepsilon(\textbf{k},\textbf{Q})}{\omega^2-\varepsilon^2(\textbf{k},\textbf{Q})},\qquad J_{a,b}=\frac{1}{N}\sum_\textbf{k}
\frac{\omega^2 a_{\textbf{k},\textbf{Q}}b_{\textbf{k},\textbf{Q}}}{\omega^2-\varepsilon^2(\textbf{k},\textbf{Q})}.
$$
Here,  $a$ and $b$ are one of the form factors (\ref{FF1}).

The above relations can be easily extended to the case of a moving lattice. In the presence of quasimomentum
$\textbf{p}$ the following definitions and relationships hold:
$$u_{\textbf{p}}(\textbf{k})=\sqrt{\frac{1}{2}\left[1+
\chi(\textbf{k};\textbf{p})/E_{\textbf{p}}(\textbf{k})\right]},\quad
v_{\textbf{p}}(\textbf{k})=\sqrt{\frac{1}{2}\left[1-
\chi(\textbf{k};\textbf{p})/E_{\textbf{p}}(\textbf{k})\right]},\quad
E{\textbf{p}}(\textbf{k})=
\sqrt{\chi^2(\textbf{k};\textbf{p})+\Delta_\textbf{p}^2},$$
$$\chi(\textbf{k};\textbf{p})=
\frac{1}{2}\left[\xi(\textbf{p}+\textbf{k})+
\xi(\textbf{k}-\textbf{p})\right],\quad \eta(\textbf{k};\textbf{p})=
\frac{1}{2}\left[\xi(\textbf{p}+\textbf{k})-
\xi(\textbf{k}-\textbf{p})\right], \quad
E_{\pm}(\textbf{k};\textbf{p})= \eta(\textbf{k};\textbf{p})\pm
E_{\textbf{p}}(\textbf{k}).$$ At zero temperature the   gap equation
for $\Delta_{\textbf{p}}$ and the particle number equation are:
$$
f=1-\frac{1}{N}\sum_\textbf{k} \frac{\chi(\textbf{k};\textbf{p})}{E{\textbf{p}}(\textbf{k})},\quad
1=\frac{|U|}{N}\sum_\textbf{k}
\frac{1}{2E{\textbf{p}}(\textbf{k})}.$$
 In the presence of quasimomentum
$\textbf{p}$  the elements of the $16\times 16$ determinant
$\widehat{Z}$ are the same as in the case of a stationary lattice,
but the definitions of $I_{a,b}$ and $J_{a,b}$ are as follows:
$$I_{a,b}=\frac{1}{N}\sum_\textbf{k}
\frac{a^\textbf{p}_{\textbf{k},\textbf{Q}}b^\textbf{p}_{\textbf{k},\textbf{Q}}\varepsilon_{\textbf{p}}(\textbf{k},\textbf{Q})}
{\left[\omega+\Omega_{\textbf{p}}(\textbf{k},\textbf{Q})\right]^2-\varepsilon^2_{\textbf{p}}(\textbf{k},\textbf{Q})},\qquad J_{a,b}=\frac{1}{N}\sum_\textbf{k}
\frac{ a^\textbf{p}_{\textbf{k},\textbf{Q}}b^\textbf{p}_{\textbf{k},\textbf{Q}}\left[\omega+\Omega_{\textbf{p}}(\textbf{k},\textbf{Q})\right]}
{\left[\omega+\Omega_{\textbf{p}}(\textbf{k},\textbf{Q})\right]^2-\varepsilon^2_{\textbf{p}}(\textbf{k},\textbf{Q})},
$$ where $\varepsilon_{\textbf{p}}(\textbf{k},\textbf{Q})=
 E_{\textbf{p}}(\textbf{k}+\textbf{Q})+ E_{\textbf{p}}(\textbf{k})$ and $\Omega_{\textbf{p}}(\textbf{k},\textbf{Q})
  =\eta_{\textbf{p}}(\textbf{k})-\eta_{\textbf{p}}
 (\textbf{k}+\textbf{Q})$. $a$ and $b$ are one of the following form factors:
$$\gamma^{\textbf{p}}_{\textbf{k},\textbf{Q}}=
u_{\textbf{p}}(\textbf{k})u_{\textbf{p}}(\textbf{k}+\textbf{Q})
+v_{\textbf{p}}(\textbf{k})v_{\textbf{p}}(\textbf{k}+\textbf{Q}),
\quad
l^{\textbf{p}}_{\textbf{k},\textbf{Q}}=u_{\textbf{p}}(\textbf{k})
u_{\textbf{p}}(\textbf{k}+\textbf{Q})
-v_{\textbf{p}}(\textbf{k})v_{\textbf{p}}(\textbf{k}+\textbf{Q}),$$
$$\widetilde{\gamma}_{\textbf{p}}{\textbf{k},\textbf{Q}}=
u_{\textbf{p}}(\textbf{k})v_{\textbf{p}}(\textbf{k}+\textbf{Q})
-v_{\textbf{p}}(\textbf{k})u_{\textbf{p}}(\textbf{k}+\textbf{Q}),\quad
 m^{\textbf{p}}_{\textbf{k},\textbf{Q}}=
u_{\textbf{p}}(\textbf{k})v_{\textbf{p}}(\textbf{k}+\textbf{Q})
+v_{\textbf{p}}(\textbf{k})u_{\textbf{p}}(\textbf{k}+\textbf{Q}).$$\end{widetext}
It is well-known that the singularity of the integrals $I_{a,b}$ and $J_{a,b}$  at energies $\omega=\varepsilon_{\textbf{p}}(\textbf{k},\textbf{Q})
-\Omega_{\textbf{p}}(\textbf{k},\textbf{Q})$  corresponds physically to the possibility of depairing into two fermion excitations with energies $E_{+}(\textbf{k}+\textbf{Q};\textbf{p})$ and $- E_{-}(\textbf{k};\textbf{p})$. The spectrum for this kind of excitation is known as the  particle-hole continuum.\cite{Com} The lower
boundary of the particle-hole continuum is defined by the condition $min_\textbf{k}\left(E_{+}(\textbf{k}+\textbf{Q};\textbf{p})- E_{-}(\textbf{k};\textbf{p})\right)$ where the minimum is to be taken over all the possible values of $\textbf{k}$.

\end{document}